\documentclass[RNAAS]{aastex631}

\begin{document}

\title{NGC 1605 is not a binary cluster}


\author[0000-0003-4524-9363]{Friedrich Anders}
\affiliation{Institut de Ci\`encies del Cosmos, Universitat de Barcelona (IEEC-UB), Mart\'i i Franqu\`es 1, 08028 Barcelona, Spain}

\author[0000-0002-9419-3725]{Alfred Castro-Ginard}
\affiliation{Leiden Observatory, Leiden University, Niels Bohrweg, 2, 2333CA Leiden, The Netherlands}

\author[0000-0003-4105-2520]{Juan Casado}
\affiliation{Facultat de Ciències, Universitat Autònoma de Barcelona (UAB), 08193 Bellaterra, Barcelona, Spain}

\author[0000-0001-5495-9602]{Carme Jordi}
\affiliation{Institut de Ci\`encies del Cosmos, Universitat de Barcelona (IEEC-UB), Mart\'i i Franqu\`es 1, 08028 Barcelona, Spain}

\author[0000-0001-9789-7069]{Lola Balaguer-Nú\~nez}
\affiliation{Institut de Ci\`encies del Cosmos, Universitat de Barcelona (IEEC-UB), Mart\'i i Franqu\`es 1, 08028 Barcelona, Spain}


\section{Abstract}
The open star cluster NGC 1605 has recently been reported to in fact consist of two clusters (one intermediate-aged and one old) that merged via a flyby capture. Here we show that {\it Gaia} data do not support this scenario. We also report the serendipitous discovery of a new open cluster, Can Batlló 1, with a similar age and distance.

\section{Introduction}
Gravitational captures of star clusters by other clusters are very rare and elusive events that can serve as laboratories for star cluster destruction (e.g. \citealt{Soubiran2018}). There are very few, if any, old binary clusters in the Galaxy \citep{Casado2022}. Recently, \citet{Piatti2022} presented a promising candidate for an open cluster collision of the nearby ($d\sim330$ pc) objects IC 4665 and Collinder 350.
Some months earlier, \citet{Camargo2021} reported the existence of a possible old binary cluster, dubbed NGC 1605a/b, and suggested that it origined from a flyby capture. The author argued that the long-known cluster NGC 1605 actually consists of two components that have vastly different ages (2 Gyr and 600 Myr). 
In the previous literature on NGC 1605 no evidence for such a duplicity was found. Here we report that both manual analysis (by each of the authors individually) and commonly used clustering analysis techniques show no hint for multiple populations in this cluster. 

\section{Gaia EDR3 analysis}
\begin{figure}
\begin{center}
\includegraphics[width=0.88\textwidth,angle=0]{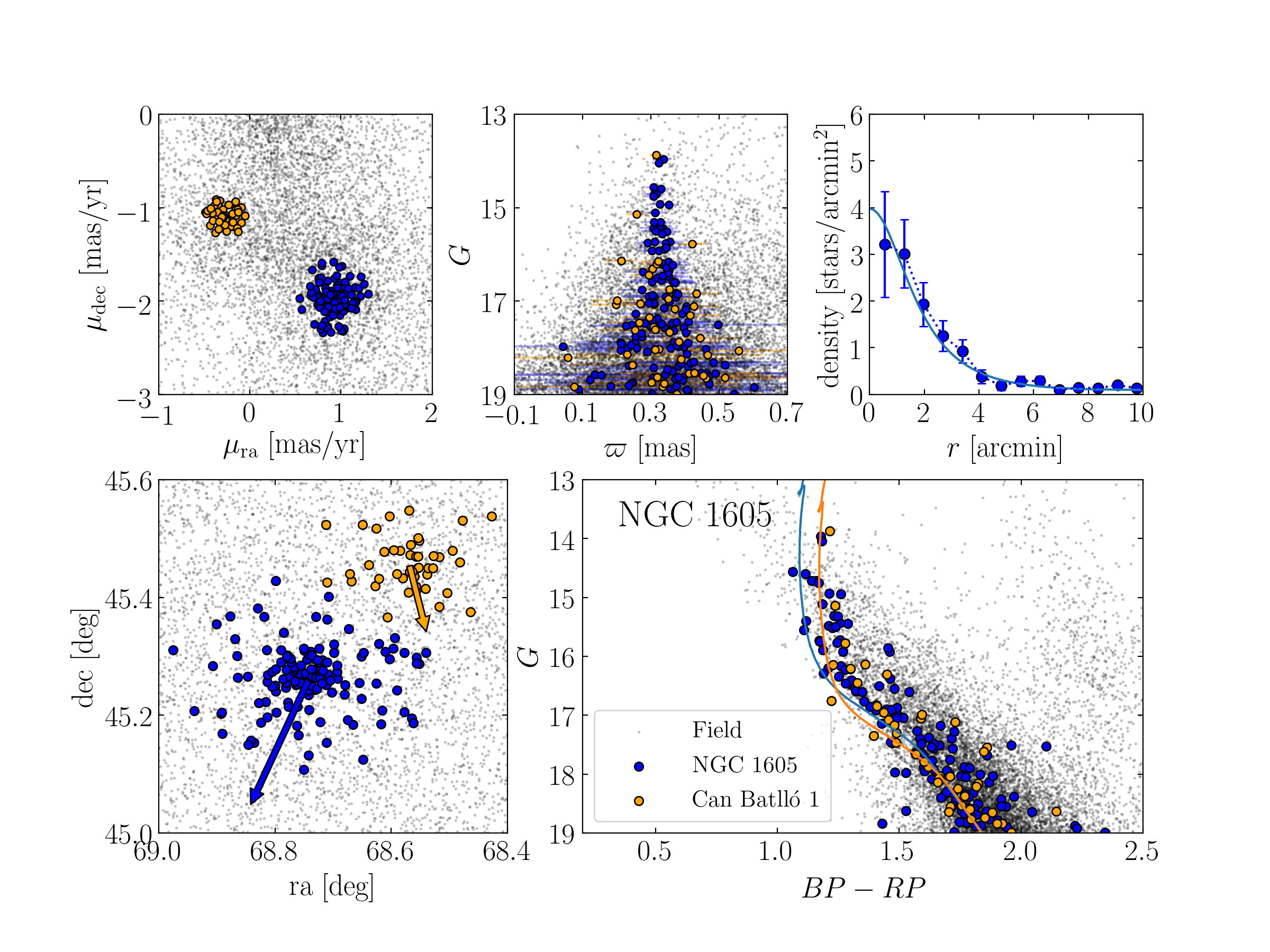}
\caption{Results of the manual analysis of the putative binary cluster NGC 1605. In each panel, members of NGC 1605 and Can Batlló 1 are highlighted in blue and orange, respectively. Bottom left panel: Sky distribution of {\it Gaia} EDR3 stars ($G<19$), with average proper motions of the two clusters indicated by arrows. Upper left panel: Proper motion diagram. Top middle: Parallax versus magnitude. Top right: density profile (including a King profile with $r_c=2$ arcmin, $r_t=10$ arcmin for comparison). Bottom right panel: Colour-magnitude diagram. Also shown are two PARSEC isochrones (blue: log age $=8.3$, [M/H]$=-0.2$, $A_V=2.6$ mag, $d=2.75$ kpc; orange: log age $=8.25$, [M/H]$=-0.4$, $A_V=2.8$ mag, $d=2.86$ kpc) . 
\label{fig:1}}
\end{center}
\end{figure}

We reanalyse the {\it Gaia} EDR3 data \citep{Gaia2021} down to magnitude $G<19$ in a 30 arcmin circle around the centre of NGC 1605 (Fig. \ref{fig:1}). We first carry out a blind analysis using the three state-of-the-art clustering techniques that have been introduced in the field: The DBSCAN algorithm employed by \citet{Castro2022}, the pyUPMASK code \citep{Pera2021}, and HDBSCAN, the preferred method of \citet{Hunt2021}. 
While the first two algorithms yield only one cluster in the considered region (NGC 1605), HDBSCAN does find another candidate close by - located about 20 arcmin northwest of NGC 1605 and clearly visible as an overdensity in proper-motion space. A manual analysis confirms these results.\footnote{Both the manual and the blind analysis are reproducible on Github \url{https://github.com/fjaellet/ngc1605} (or via the archived version at \url{https://doi.org/10.5281/zenodo.6353880}).}

NGC 1605's stellar density on the sky looks slightly irregular at first sight (it seems to be missing stars in its centre). It is, however, at all radii 2$\sigma$-consistent with a typical \citet{King1962} profile. There are also no irregularities in proper motion or in parallax space. We find no evidence for the tidal streams claimed by \citet{Camargo2021} - their claimed location would also be dynamically inconsistent with the proper motion of the putative sub-clusters.
The second sequence that \citet{Camargo2021} found in the infra-red colour-magnitude diagram of the region (their Fig. 6) is likely produced by poorly removed field-star contamination (see Sect. 4 of \citealt{CantatAnders2020} for a discussion).

In summary, we caution against purely visual photometric analysis of star clusters when {\it Gaia} data are available. We find no evidence for NGC 1605 being the old binary cluster advertised by \citet{Camargo2021}. Nevertheless, NGC 1605 is an intriguing object: Its large Galactocentric distance ($\sim 11$ kpc) and the newly discovered nearby object, Can Batlló 1, of similar age and possibly less than 100 pc away from NGC 1605, make it an interesting target for follow-up observations. Correcting for the {\it Gaia} EDR3 parallax zeropoint offsets \citep{Lindegren2021}, we obtain Bayesian distances of $2.75\pm 0.04$ kpc for NGC 1605 and $2.86\pm0.09$ kpc for Can Batlló 1. Tentative estimates of the cluster parameters are given in the caption of Fig. 1 and justified in the online material.

\vspace{-0.2cm}
\bibliography{ngc1605}{}

\begin{thebibliography}{}
\expandafter\ifx\csname natexlab\endcsname\relax\def\natexlab#1{#1}\fi
\providecommand{\url}[1]{\href{#1}{#1}}
\providecommand{\dodoi}[1]{doi:~\href{http://doi.org/#1}{\nolinkurl{#1}}}
\providecommand{\doeprint}[1]{\href{http://ascl.net/#1}{\nolinkurl{http://ascl.net/#1}}}
\providecommand{\doarXiv}[1]{\href{https://arxiv.org/abs/#1}{\nolinkurl{https://arxiv.org/abs/#1}}}

\bibitem[{{Camargo}(2021)}]{Camargo2021}
{Camargo}, D. 2021, \apj, 923, 21, \dodoi{10.3847/1538-4357/ac2835}

\bibitem[{{Cantat-Gaudin} \& {Anders}(2020)}]{CantatAnders2020}
{Cantat-Gaudin}, T., \& {Anders}, F. 2020, \aap, 633, A99,
  \dodoi{10.1051/0004-6361/201936691}

\bibitem[{{Casado}(2022)}]{Casado2022}
{Casado}, J. 2022, Universe, 8, 113, \dodoi{10.3390/universe8020113}

\bibitem[{{Castro-Ginard} {et~al.}(2022){Castro-Ginard}, {Jordi}, {Luri},
  {Cantat-Gaudin}, {Carrasco}, {Casamiquela}, {Anders},
  {Balaguer-N{\'u}{\~n}ez}, \& {Badia}}]{Castro2022}
{Castro-Ginard}, A., {Jordi}, C., {Luri}, X., {et~al.} 2022, \aap, accepted.
\newblock \doarXiv{2111.01819}

\bibitem[{{Gaia Collaboration} {et~al.}(2021){Gaia Collaboration}, {Brown},
  {Vallenari}, {Prusti}, {de Bruijne}, {Babusiaux}, {Biermann}, {Creevey},
  {Evans}, {Eyer}, {Hutton}, {Jansen}, {Jordi}, {Klioner}, {Lammers},
  {Lindegren}, {Luri}, {Mignard}, {Panem}, {Pourbaix}, {Randich}, {Sartoretti},
  {Soubiran}, {Walton}, {Arenou}, {Bailer-Jones}, {Bastian}, {Cropper},
  {Drimmel}, {Katz}, {Lattanzi}, {van Leeuwen}, {Bakker}, {Cacciari},
  {Casta{\~n}eda}, {De Angeli}, {Ducourant}, {Fabricius}, {Fouesneau},
  {Fr{\'e}mat}, {Guerra}, {Guerrier}, {Guiraud}, {Jean-Antoine Piccolo},
  {Masana}, {Messineo}, {Mowlavi}, {Nicolas}, {Nienartowicz}, {Pailler},
  {Panuzzo}, {Riclet}, {Roux}, {Seabroke}, {Sordo}, {Tanga}, {Th{\'e}venin},
  {Gracia-Abril}, {Portell}, {Teyssier}, {Altmann}, {Andrae}, {Bellas-Velidis},
  {Benson}, {Berthier}, {Blomme}, {Brugaletta}, {Burgess}, {Busso}, {Carry},
  {Cellino}, {Cheek}, {Clementini}, {Damerdji}, {Davidson}, {Delchambre},
  {Dell'Oro}, {Fern{\'a}ndez-Hern{\'a}ndez}, {Galluccio}, {Garc{\'\i}a-Lario},
  {Garcia-Reinaldos}, {Gonz{\'a}lez-N{\'u}{\~n}ez}, {Gosset}, {Haigron},
  {Halbwachs}, {Hambly}, {Harrison}, {Hatzidimitriou}, {Heiter},
  {Hern{\'a}ndez}, {Hestroffer}, {Hodgkin}, {Holl}, {Jan{\ss}en}, {Jevardat de
  Fombelle}, {Jordan}, {Krone-Martins}, {Lanzafame}, {L{\"o}ffler}, {Lorca},
  {Manteiga}, {Marchal}, {Marrese}, {Moitinho}, {Mora}, {Muinonen}, {Osborne},
  {Pancino}, {Pauwels}, {Petit}, {Recio-Blanco}, {Richards}, {Riello},
  {Rimoldini}, {Robin}, {Roegiers}, {Rybizki}, {Sarro}, {Siopis}, {Smith},
  {Sozzetti}, {Ulla}, {Utrilla}, {van Leeuwen}, {van Reeven}, {Abbas}, {Abreu
  Aramburu}, {Accart}, {Aerts}, {Aguado}, {Ajaj}, {Altavilla}, {{\'A}lvarez},
  {{\'A}lvarez Cid-Fuentes}, {Alves}, {Anderson}, {Anglada Varela}, {Antoja},
  {Audard}, {Baines}, {Baker}, {Balaguer-N{\'u}{\~n}ez}, {Balbinot}, {Balog},
  {Barache}, {Barbato}, {Barros}, {Barstow}, {Bartolom{\'e}}, {Bassilana},
  {Bauchet}, {Baudesson-Stella}, {Becciani}, {Bellazzini}, {Bernet}, {Bertone},
  {Bianchi}, {Blanco-Cuaresma}, {Boch}, {Bombrun}, {Bossini}, {Bouquillon},
  {Bragaglia}, {Bramante}, {Breedt}, {Bressan}, {Brouillet}, {Bucciarelli},
  {Burlacu}, {Busonero}, {Butkevich}, {Buzzi}, {Caffau}, {Cancelliere},
  {C{\'a}novas}, {Cantat-Gaudin}, {Carballo}, {Carlucci}, {Carnerero},
  {Carrasco}, {Casamiquela}, {Castellani}, {Castro-Ginard}, {Castro Sampol},
  {Chaoul}, {Charlot}, {Chemin}, {Chiavassa}, {Cioni}, {Comoretto}, {Cooper},
  {Cornez}, {Cowell}, {Crifo}, {Crosta}, {Crowley}, {Dafonte}, {Dapergolas},
  {David}, {David}, {de Laverny}, {De Luise}, {De March}, {De Ridder}, {de
  Souza}, {de Teodoro}, {de Torres}, {del Peloso}, {del Pozo}, {Delbo},
  {Delgado}, {Delgado}, {Delisle}, {Di Matteo}, {Diakite}, {Diener},
  {Distefano}, {Dolding}, {Eappachen}, {Edvardsson}, {Enke}, {Esquej}, {Fabre},
  {Fabrizio}, {Faigler}, {Fedorets}, {Fernique}, {Fienga}, {Figueras},
  {Fouron}, {Fragkoudi}, {Fraile}, {Franke}, {Gai}, {Garabato},
  {Garcia-Gutierrez}, {Garc{\'\i}a-Torres}, {Garofalo}, {Gavras}, {Gerlach},
  {Geyer}, {Giacobbe}, {Gilmore}, {Girona}, {Giuffrida}, {Gomel}, {Gomez},
  {Gonzalez-Santamaria}, {Gonz{\'a}lez-Vidal}, {Granvik},
  {Guti{\'e}rrez-S{\'a}nchez}, {Guy}, {Hauser}, {Haywood}, {Helmi}, {Hidalgo},
  {Hilger}, {H{\l}adczuk}, {Hobbs}, {Holland}, {Huckle}, {Jasniewicz},
  {Jonker}, {Juaristi Campillo}, {Julbe}, {Karbevska}, {Kervella}, {Khanna},
  {Kochoska}, {Kontizas}, {Kordopatis}, {Korn}, {Kostrzewa-Rutkowska},
  {Kruszy{\'n}ska}, {Lambert}, {Lanza}, {Lasne}, {Le Campion}, {Le Fustec},
  {Lebreton}, {Lebzelter}, {Leccia}, {Leclerc}, {Lecoeur-Taibi}, {Liao},
  {Licata}, {Lindstr{\o}m}, {Lister}, {Livanou}, {Lobel}, {Madrero Pardo},
  {Managau}, {Mann}, {Marchant}, {Marconi}, {Marcos Santos}, {Marinoni},
  {Marocco}, {Marshall}, {Martin Polo}, {Mart{\'\i}n-Fleitas}, {Masip},
  {Massari}, {Mastrobuono-Battisti}, {Mazeh}, {McMillan}, {Messina},
  {Michalik}, {Millar}, {Mints}, {Molina}, {Molinaro}, {Moln{\'a}r},
  {Montegriffo}, {Mor}, {Morbidelli}, {Morel}, {Morris}, {Mulone}, {Munoz},
  {Muraveva}, {Murphy}, {Musella}, {Noval}, {Ord{\'e}novic}, {Orr{\`u}},
  {Osinde}, {Pagani}, {Pagano}, {Palaversa}, {Palicio}, {Panahi}, {Pawlak},
  {Pe{\~n}alosa Esteller}, {Penttil{\"a}}, {Piersimoni}, {Pineau}, {Plachy},
  {Plum}, {Poggio}, {Poretti}, {Poujoulet}, {Pr{\v{s}}a}, {Pulone}, {Racero},
  {Ragaini}, {Rainer}, {Raiteri}, {Rambaux}, {Ramos}, {Ramos-Lerate}, {Re
  Fiorentin}, {Regibo}, {Reyl{\'e}}, {Ripepi}, {Riva}, {Rixon}, {Robichon},
  {Robin}, {Roelens}, {Rohrbasser}, {Romero-G{\'o}mez}, {Rowell}, {Royer},
  {Rybicki}, {Sadowski}, {Sagrist{\`a} Sell{\'e}s}, {Sahlmann}, {Salgado},
  {Salguero}, {Samaras}, {Sanchez Gimenez}, {Sanna}, {Santove{\~n}a},
  {Sarasso}, {Schultheis}, {Sciacca}, {Segol}, {Segovia}, {S{\'e}gransan},
  {Semeux}, {Shahaf}, {Siddiqui}, {Siebert}, {Siltala}, {Slezak}, {Smart},
  {Solano}, {Solitro}, {Souami}, {Souchay}, {Spagna}, {Spoto}, {Steele},
  {Steidelm{\"u}ller}, {Stephenson}, {S{\"u}veges}, {Szabados}, {Szegedi-Elek},
  {Taris}, {Tauran}, {Taylor}, {Teixeira}, {Thuillot}, {Tonello}, {Torra},
  {Torra}, {Turon}, {Unger}, {Vaillant}, {van Dillen}, {Vanel}, {Vecchiato},
  {Viala}, {Vicente}, {Voutsinas}, {Weiler}, {Wevers}, {Wyrzykowski}, {Yoldas},
  {Yvard}, {Zhao}, {Zorec}, {Zucker}, {Zurbach}, \& {Zwitter}}]{Gaia2021}
{Gaia Collaboration}, {Brown}, A.~G.~A., {Vallenari}, A., {et~al.} 2021, \aap,
  649, A1, \dodoi{10.1051/0004-6361/202039657}

\bibitem[{{Hunt} \& {Reffert}(2021)}]{Hunt2021}
{Hunt}, E.~L., \& {Reffert}, S. 2021, \aap, 646, A104,
  \dodoi{10.1051/0004-6361/202039341}

\bibitem[{{King}(1962)}]{King1962}
{King}, I. 1962, \aj, 67, 471, \dodoi{10.1086/108756}

\bibitem[{{Lindegren} {et~al.}(2021){Lindegren}, {Bastian}, {Biermann},
  {Bombrun}, {de Torres}, {Gerlach}, {Geyer}, {Hern{\'a}ndez}, {Hilger},
  {Hobbs}, {Klioner}, {Lammers}, {McMillan}, {Ramos-Lerate},
  {Steidelm{\"u}ller}, {Stephenson}, \& {van Leeuwen}}]{Lindegren2021}
{Lindegren}, L., {Bastian}, U., {Biermann}, M., {et~al.} 2021, \aap, 649, A4,
  \dodoi{10.1051/0004-6361/202039653}

\bibitem[{{Pera} {et~al.}(2021){Pera}, {Perren}, {Moitinho}, {Navone},
  {Vazquez}, {some}, \& {author}}]{Pera2021}
{Pera}, M.~S., {Perren}, G.~I., {Moitinho}, A., {et~al.} 2021, \aap, 650, A109,
  \dodoi{10.1051/0004-6361/202040252}

\bibitem[{{Piatti} \& {Malhan}(2022)}]{Piatti2022}
{Piatti}, A.~E., \& {Malhan}, K. 2022, \mnras, 511, L1,
  \dodoi{10.1093/mnrasl/slab130}

\bibitem[{{Soubiran} {et~al.}(2018){Soubiran}, {Cantat-Gaudin},
  {Romero-G{\'o}mez}, {Casamiquela}, {Jordi}, {Vallenari}, {Antoja},
  {Balaguer-N{\'u}{\~n}ez}, {Bossini}, {Bragaglia}, {Carrera}, {Castro-Ginard},
  {Figueras}, {Heiter}, {Katz}, {Krone-Martins}, {Le Campion}, {Moitinho}, \&
  {Sordo}}]{Soubiran2018}
{Soubiran}, C., {Cantat-Gaudin}, T., {Romero-G{\'o}mez}, M., {et~al.} 2018,
  \aap, 619, A155, \dodoi{10.1051/0004-6361/201834020}

\end{thebibliography}
\bibliographystyle{aasjournal}

\end{document}